\documentclass[aps,prc,twocolumn,nofootinbib,floatfix,amsmath,amssymb,amsfonts,superscriptaddress]{revtex4-2}

\usepackage{times}
\usepackage{txfonts}

\usepackage{graphicx} 
\usepackage{dcolumn}  
\usepackage{xcolor}
\usepackage[
]{hyperref}
\usepackage[nameinlink]{cleveref}
\crefname{table}{Table}{Tables}
\Crefname{table}{Table}{Tables}
\crefname{figure}{Fig.}{Figs.}
\Crefname{figure}{Figure}{Figures}
\crefname{section}{Sec.}{Secs.}
\Crefname{section}{Section}{Sections}
\crefname{equation}{Eq.}{Eqs.}
\Crefname{equation}{Equation}{Equations}

\usepackage[
    range-units=single
    ]{siunitx}
\usepackage{dcolumn}
\newcolumntype{d}[1]{D{.}{.}{#1}}

\usepackage[caption=false]{subfig}

\graphicspath{{./}}

\usepackage{isotope}
\newcommand*{\elem}[2]{\ensuremath{\isotope[#2]{\mathrm{#1}}}}

\DeclareMathAlphabet{\curly}{OMS}{cmsy}{m}{n}

\begin{document}

\title{Benchmarking ANN extrapolations of the ground-state energies and radii of Li isotopes}

\author{M.~Kn\"oll}
\email{marco.knoell@physik.tu-darmstadt.de}
\affiliation{Institut f\"ur Kernphysik, Technische Universit\"at Darmstadt, 64289 Darmstadt, Germany}
\author{M.~Lockner} 
\affiliation{Dept.\ of Physics and Astronomy, Iowa State University, Ames, Iowa 50011, USA}
\author{P.~Maris} 
\email{pmaris@iastate.edu}
\affiliation{Dept.\ of Physics and Astronomy, Iowa State University, Ames, Iowa 50011, USA}
\author{R.J.~McCarty} 
\affiliation{Dept.\ of Physics and Astronomy, Iowa State University, Ames, Iowa 50011, USA}
\author{R.~Roth}
\email{robert.roth@physik.tu-darmstadt.de}
\affiliation{Institut f\"ur Kernphysik, Technische Universit\"at Darmstadt, 64289 Darmstadt, Germany}
\affiliation{Helmholtz Forschungsakademie Hessen f\"ur FAIR, GSI Helmholtzzentrum, 64289 Darmstadt, Germany}
\author{J.P.~Vary}
\email{jvary@iastate.edu}
\affiliation{Dept.\ of Physics and Astronomy, Iowa State University, Ames, Iowa 50011, USA}
\author{T.~Wolfgruber}
\affiliation{Institut f\"ur Kernphysik, Technische Universit\"at Darmstadt, 64289 Darmstadt, Germany}

\date{\today}

\begin{abstract}
We present a comparison of model-space extrapolation methods for No-Core Shell Model calculations of ground-state energies and root-mean-square radii in Li isotopes.
In particular, we benchmark the latest machine learning tools against widely used exponential and infrared extrapolations for energies and crossing point estimates for radii.
Our findings demonstrate that while some machine learning-based approaches provide reliable predictions with robust statistical uncertainties for both observables even in small model spaces, others perform less consistently.
The predictions of the former are compatible with established exponential and IR extrapolations for energies while also providing a notable improvement over conventional radius estimates.

\end{abstract}

\maketitle

\section{Introduction}

One of the major challenges in nuclear structure theory is the solution of the nuclear many-body problem.
With the advent of high-performance computers, ab initio configuration interaction methods such as the No-Core Shell Model (NCSM)~\cite{NaQu09,Barrett:2013nh} have emerged as a powerful tool to compute a variety of nuclear properties from realistic Hamiltonians \cite{Hergert:2020}.
However, rapidly growing model spaces, especially with particle number, severely limit their applicability and convergence can only be reached for selected few-body systems.
Even more sophisticated model-space truncation schemes, e.g., importance truncation \cite{Roth09}, or improved single-particle wavefunctions~\cite{Caprio:2012rv,Tichai2019Natural}, deliver only minor gains in reach of these methods. 

To this day, the extrapolation from truncated model spaces to the full Hilbert space remains a formidable task and different attempts to emulate the model-space dependence have been developed.
Traditionally, these attempts are based on exponential or polynomial fits to the results obtained in small model spaces and provide good approximations to the ground- and excited-state energies \cite{Bogner:2007rx,Maris:2008ax,Roth09,JuMa13,LENPIC21}.
However, the quality of these extrapolations decreases considerably when applied to other observables such as radii, which, unlike energies, do not obey a variational principle.

A second generation of physics-motivated extrapolation schemes has been developed with regard to the short and long-range parts of the associated wavefunction. 
Given the assumption that the short-range part is converged, the missing long-range contributions can be modeled using an exponential function \cite{Furnstahl:2012ir,Coon:2012ir,More:2013ir,Furnstahl:2015ir,Wendt:2015ir}.
Although these so-called infrared (IR) extrapolations can be successfully applied to energies and extensions to radii have been explored, it introduces additional constraints on the many-body calculation.

More recently, Machine-Learning (ML) approaches have been employed to address the model-space extrapolation problem \cite{Negoita:2018yok,Negoita:2018kgi,JiHa19,KnoWo23,Wolfgruber:2023ehw,mazur2024machine}.
Based on their capabilities as universal function approximators, Artificial Neural Networks (ANNs) are the ideal tool to emulate the unknown model-space dependence and different ML approaches yield high-quality extrapolations along with uncertainty estimates for energies and radii.

In this work, we provide a detailed benchmark for the ML extrapolation schemes developed at Iowa State University (ISU) \cite{Negoita:2018kgi,Negoita:2018yok} and at TU Darmstadt (TUDa) \cite{KnoWo23,Wolfgruber:2023ehw} for energies and radii in Li isotopes.
We further discuss how they compare to exponential and IR extrapolation schemes.

\section{Methods}

\begin{figure*}
    \centering%
    \includegraphics[width=\linewidth]{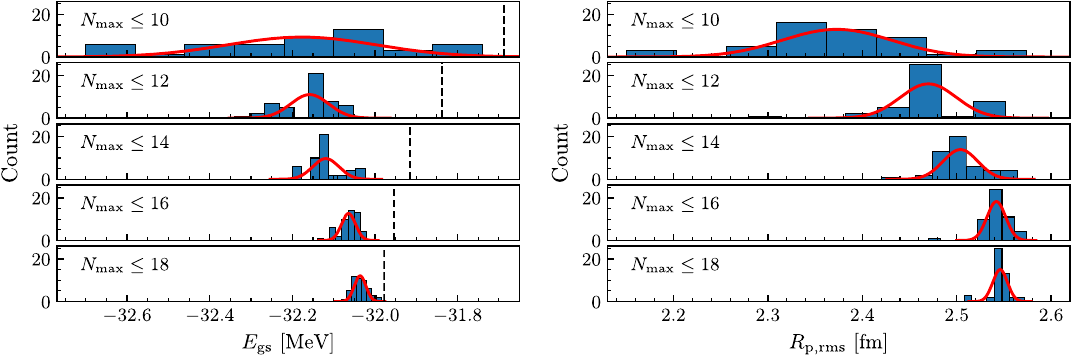}
    \caption{A selection of the ISU $^6$Li ground-state energy (left-hand panels) and point-proton radius (right-hand panels) histograms. A Gaussian fit has been applied to our binned histograms with the mean value of our fit taken to be the value of our observable and the standard deviation taken as our error bars. The count is the total number of networks represented in that binning.  The black dashed vertical lines in the energy histograms indicate the variational minimum at each respective upper limit $N_{\rm max}$ value.}
    \label{ISU_Li6}
\end{figure*}

In the NCSM approach, the wavefunction $\Psi$ of a nucleus consisting of $Z$ protons and $N$ neutrons is expanded in an $A=(Z+N)$-body basis of Slater determinants of Harmonic Oscillator (HO) single-particle wavefunctions.   In a HO basis, the single-particle wavefunctions $\phi_{nljm}(\vec{r})$ are labeled by their radial quantum number $n$, orbital motion quantum number $l$, total single-particle spin $j=l\pm \frac{1}{2}$, and magnetic projection $m$ which satisfies $-j \le m \le j$; in addition, the HO basis functions depend on the HO basis scale $\hbar\Omega$.  Furthermore, if we apply a truncation on the total number of HO quanta in the system, that is, a truncation on $\sum_A (2n+l)$ over all $A$ nucleons, the obtained wavefunctions $\Psi$ factorize exactly into center-of-mass wavefunctions and relative wavefunctions~\cite{Lipkin:1958zz,Caprio:2020heu}; and typically, we use a Lagrange multiplier to remove states with center-of-mass excitations from the low-lying spectrum.  We use the notation $N_{\rm max}$ as our basis truncation parameter, which is defined as the total number of HO oscillator quanta above the valence configuration.  Because of parity conservation, $N_{\rm max}$ is increased in steps of two, starting from $N_{\rm max}=0$.

For any finite basis expansion, the obtained energy $E$ gives a strict upper bound for the energy in the complete, but infinite-dimensional basis, at least for the lowest states of a given $Z$, $N$, and spin-parity quantum numbers $J^P$; and the corresponding eigenvector $\vec{c}$ gives an approximation to the $A$-body wavefunction $\Psi(\vec{r}_1, \ldots, \vec{r}_A)$.  As one increases the basis size, the obtained eigenvectors approach the exact $A$-body wavefunction for a given Hamiltonian.  The challenge is to extract converged values for physical observables based on a series of calculations in finite model spaces.

Here we consider the ground-state energies and root-mean-square (rms) point-proton radii of $^6$Li, $^7$Li, and $^8$Li, calculated within the NCSM using a realistic nucleon-nucleon potential.  For these three nuclei, we have performed NCSM calculations over a range of $\hbar\Omega$ values from $10$~MeV up to $30$~MeV in steps of $2.5$~MeV.  For our extrapolations to the complete (infinite-dimensional) basis we consider NCSM results obtained with $N_{\rm max} = 2$ up through $N_{\rm max} = 18$ for $^6$Li, up through $N_{\rm max} = 16$ for $^7$Li, and $N_{\rm max} = 14$ for $^8$Li.  

All NCSM calculations presented here were performed with the code Many-Fermion Dynamics nuclear~\cite{MARIS201097,doi:10.1002/cpe.3129,SHAO20181,MARIS2022101554} using the Daejeon16 nucleon-nucleon potential~\cite{Shirokov:2016ead} at the National Energy Research Scientific Computing center (NERSC).  Daejeon16 is a two-body potential based on a chiral effective field theory potential at N$^3$LO~\cite{Entem:2001cg,Entem:2003ft}, softened to improve numerical convergence for NCSM calculations by Similarity Renormalization Group (SRG) evolution to a scale of $\lambda = 1.5$~fm$^{-1}$.  In addition, a set of Phase Equivalent Transformations (PETs) was used to fit the energy levels of 11 $s$- and $p$-shell states.  The resulting Daejeon16 potential gives an excellent description of the low-lying spectra of all $p$-shell nuclei~\cite{Maris:2019etr}.

\subsection{ISU ANNs}

The ISU ANNs are based on the pioneering work of Refs.~\cite{Negoita:2018kgi,Negoita:2018yok}.  
For each observable, we construct ANNs with an input layer that takes a table of results that have been calculated in the NCSM at different $\hbar\Omega$ and $N_{\rm max}$ values.
These inputs are then sent through a hidden layer consisting of eight nodes using a feed-forward algorithm.  
Each network is trained through this layer a total of 10 times, using 85\% of the input data, and tested with the remaining 15\% of the input data.
We use the rms error on these test data to select the best performing networks out of a large ensemble of trained networks in order to predict the behavior of our desired observable at a large value of $N_{\rm max}$.  
After selecting on the rms error, we also apply a test on the variation of the predicted observable over the chosen $\hbar\Omega$ range at large $N_{\rm max}$ \cite{McCarty:2024diss}, and for the ground-state energies, a test on (approximate) monotonic decreasing behavior with increasing $N_{\rm max}$ so that the variational principle is (approximately) satisfied \cite{Negoita:2018kgi,McCarty:2024diss}.
Finally, we remove outliers with a three-sigma rule \cite{Negoita:2018kgi}.

For each observable we consider up to five input data sets, each over the full range of $\hbar\Omega$ from 10 to 30 MeV, but with $N_{\rm max}$ increasing from $N_{\rm max} = 10$ up to $18$ for $^6$Li, up to $N_{\rm max} = 16$ for $^7$Li, and up to  $N_{\rm max} = 14$ for $^8$Li.  
Each of these input data sets is used to train and test a total of 200,000 networks, out of which we select the top 50 performing ANNs.
Note that these input data sets have an increasing number of data points, from 45 data points for $N_{\rm max} = 10$ up to at most 81 data points for $^6$Li at  $N_{\rm max} = 18$.  
Our predictions for the converged values are made at $N_{\rm max} = 70$ for the ground-state energy, and at $N_{\rm max} = 90$ for the rms point-proton radius.  
For the maximum acceptable rms error we choose $0.014$~MeV for energies, and $0.0007$~fm for radii, and for the flatness test at  $N_{\rm max} = 70 (90)$ we use a cut-off of $0.5$\% variation for the ground-state energies and 4\% for the point-proton radii over the $\hbar\Omega$ range of 10 to 30 MeV.  
In addition, for the energies we remove ANNs that violate the monotonic decreasing behavior by more than $25$ keV at any $\Delta N_{\rm max} = 2$ step for $N_{\rm max} \leq 20$ and any $\Delta N_{\rm max} = 5$ step for $N_{\rm max} > 20$.

Figure \ref{ISU_Li6} shows histograms for our top performing networks for both the ground-state energy (left) and the point-proton radius (right) of $^6$Li at each respective upper limit $N_{\rm max}$ cutoff, starting from $N_{\rm max} = 10$.  
We fit a Gaussian curve to our results and take the mean as our predicted value and the standard deviation as our error bars.  
Both for the energies and for the radii, we observe a systematic decrease in the standard deviation of our histograms with respect to increasing $N_{\rm max}$, as expected.  
Minor variations on the detailed procedure to train and select the ANNs are presented in Ref.~\cite{McCarty:2024diss}.

\subsection{TUDa ANNs}
The TUDa ANNs developed in Ref.~\cite{KnoWo23,Wolfgruber:2023ehw,Knoell2024Diss} present a different machine learning approach to the extrapolation of NCSM calculations.
Instead of emulating the functional form of observables in terms of $\hbar\Omega$ and $N_\mathrm{max}$ the converged value is directly predicted from a set of calculations in small model spaces.
This exploits the pattern recognition capabilities of ANNs, which are designed to capture the observable-specific convergence patterns in few-body systems.
The universality of these convergence patterns then allows for a subsequent application to a broad range of $p$-shell nuclei.

In particular, the converged value is predicted from sequences of four NCSM data points at consecutive $N_\mathrm{max}$ for three different $\hbar\Omega$.
As this extends beyond the approximation of a continuous function, a larger network topology is employed consisting of five layers with $12,48,48,24,1$ nodes each.

The training of the TUDa ANNs requires fully converged NCSM data, which are only accessible for systems with $A\leq 4$.
Hence, the training data consists of calculations for \elem{H}{2},\elem{H}{3}, and \elem{He}{4} with non-local chiral interactions from \cite{EnMa17,HuVo20} for a wide range of orders, cutoffs, and SRG flow parameters each for a $\hbar\Omega$ window from \qtyrange{12}{32}{MeV}.
For the actual training all possible input samples, i.e.,\ subsets of twelve data points that match the input layer of the ANN, are constructed and normalized to the interval $[0,1]$.
The networks employed in this work are the same ones used in \cite{Wolfgruber:2023ehw} and details on sampling and training are given there.

Once trained, the TUDa ANNs can be evaluated with NCSM data for any $p$-shell nucleus, interaction, and state.
At this point the extrapolation problem has essentially been converted into an interpolation problem \cite{Knoell2024Diss}.
Before being passed through the networks, the evaluation data needs to undergo the same sampling procedure as the training data.
This naturally yields several predictions for the converged value from a single ANN.
A total of 1000 ANNs are evaluated with the same data in order to incorporate any network uncertainties.
The wealth of predictions is then collected in a histogram using the Freedman-Diaconis rule \cite{Freedman:1981hist} to estimate the bin size.
From this histogram an ensemble prediction along with an uncertainty can be extracted through statistical means.
In order to avoid minor variations in bin height the histogram is first smoothed with a moving average with width 10.
Next, the ensemble prediction is given by the most probable value.
Finally, we can assign an uncertainty by adding up bins to the left and right of the most probable predictions, always continuing with the higher of the two, until the accumulated counts exceed 68\% of the total number of predictions.
Thus, the resulting interval allows for asymmetric uncertainties and can be interpreted as 68\% confidence interval.

In practical applications, the shape of the histogram depends on the selection of evaluation data.
While the ensemble prediction is rather robust under changes in the data, the inclusion of more exotic $\hbar\Omega$ typically yields larger uncertainties.
We, therefore, propose the following selection procedures and demonstrate their effectiveness in the result section of this work. 
For energies the ANNs perform best if the evaluation data is evenly distributed around the variational optimum.
Hence, we select sequences for five values of $\hbar\Omega$ such that at least one sequence is on either side of the optimum. 
For radii the selection criteria regard the shape of the converging sequences as small frequencies typically yield oscillating behavior or very slow convergence that impede the ANNs performance.
Therefore, the five smallest values of $\hbar\Omega$ which generate monotonously ascending sequences are selected for predictions. 
\begin{figure}
    \centering
    \includegraphics[width=1.\linewidth]{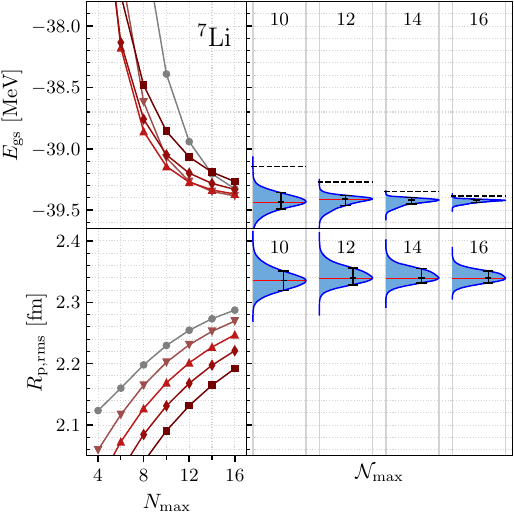}
    \caption{Evaluation of the TUDa ANNs for $^7$Li with 5 selected HO frequencies for the ground-state energy (upper panels) and the point-proton rms radius (lower panels). The left panels display the input data fed into the networks, while the right panels present histograms of the network predictions along with the extracted values and uncertainties for increasing $\curly{N}_\text{max}$, including only input data with $N_\text{max}\le \curly{N}_\text{max}$.}\label{TUDa_ANNs}
\end{figure}
Figure \ref{TUDa_ANNs} illustrates the performance of the TUDa ANNs for the ground-state energy (upper panels) and rms proton radius (lower panels) of \elem{Li}{7}.
Based on the NCSM data in the left panels, distributions of predictions at different $N_\mathrm{max}$ are obtained and depicted in the right panels.
The errorbars indicate the most probable value along with the uncertainty interval, which is additionally illustrated in orange.
The comparison of the predictions at different $N_\mathrm{max}$ allows for an assessment of the consistency and, therefore, quality of the predictions which typically exhibit decreasing uncertainties in larger model spaces.

\subsection{Empirical Exponential Extrapolations}
For comparison, we also use an empirical exponential extrapolation~\cite{Maris:2008ax} for the ground state energy
\begin{eqnarray}
  E^{\hbar\Omega}(N_{\rm max}) &=& E_\infty + c_1 \, {\rm e}^{(-c_2 N_{\rm max})} \,.
  \label{Eq:expfit}
\end{eqnarray}
At fixed values of the HO scale $\hbar\Omega$ this is a monotonically decreasing function of $N_{\rm max}$, consistent with the variational principle, which we can fit to three (or more) values of $N_{\rm max}$.  The fit parameters $c_1$ and $c_2$ depend on $\hbar\Omega$ (and $N_{\rm max}$), but the extrapolated energy $E_\infty$ should (ideally) be independent of these model space parameters.  In practice, this method gives an extrapolated energy $E_\infty$ that does depend on $N_{\rm max}$ and $\hbar\Omega$, but only weakly; and the parameters $c_1$ and $c_2$ also depend only weakly on $N_{\rm max}$; furthermore, this dependence is in general minimal around the variational minimum in $\hbar\Omega$ at fixed $N_{\rm max}$.  We use this residual dependence on $N_{\rm max}$ and $\hbar\Omega$ to make an estimate of the uncertainties in this extrapolation method.

Specifically, we use (\ref{Eq:expfit}) to fit three consecutive values of $N_{\rm max}$ at several $\hbar\Omega$ values around the variational minimum.  Next, we take as our best estimate for the converged energy $E_{\rm gs}$ the value of $E^{\hbar\Omega}_\infty$ for which $|E^{\hbar\Omega}_\infty - E^{\hbar\Omega}(N_{\rm nmax})|$ is minimal, which typically occurs for $\hbar\Omega$ at or slightly above the variational minimum.  Finally, for the uncertainty estimate we take the rms sum of the half the variation in $\hbar\Omega$ over a $7.5$~MeV window and the difference between the extrapolated value and the extrapolated value at the next-smaller $N_{\rm max}$.  This is the same method as applied in Ref.~\cite{Maris:2019etr}, and appears to work reasonably well for a range of interactions and light nuclei.

For additional comparison, we consider IR extrapolations of ground-state energies \cite{Furnstahl:2012ir,Furnstahl:2015ir,Wendt:2015ir}.
This extrapolation scheme is motivated by the different ultraviolet (UV) and IR length scales in the HO basis and the convergence w.r.t.\ these scales.
Specifically, $N_\mathrm{max}$ and $\hbar\Omega$ are converted into an UV length scale $\Lambda_\mathrm{eff}$ and an IR length scale $L_\mathrm{eff}$.
For sufficiently large values of $\Lambda_\mathrm{eff}$ the NCSM calculation is considered UV converged and the remaining dependence on $L_\mathrm{eff}$ is exponential and can, therefore, be modeled as
\begin{eqnarray}
    E(L_\mathrm{eff}) = E_\infty + c_1 e^{(-2 c_2 L_\mathrm{eff})}
\end{eqnarray}
with fit parameters $c_1$ and $c_2$.
In order to achieve UV convergence in the model spaces discussed here, large values of $\hbar\Omega$ are required. 
Thus, we consider NCSM calculation up to $\hbar\Omega = 50\,\text{MeV}$ for the IR extrapolation.
Note that these calculations are uncommonly far from the variational minimum and obtaining results for a wide range of $\hbar\Omega$ at high $N_\mathrm{max}$ is computationally expensive.
In this work, we discuss two different IR extrapolations, where the first one na{\"i}vely follows \cite{Wendt:2015ir} and only data with $\Lambda_\mathrm{eff}>800$~MeV is considered for the fit. The second one includes an additional manual selection of data, such that it falls tightly on an exponential envelope, for optimal results from the exponential fit.
For this we limit the data for the fit to the 7 to 9 data pairs with the largest $L_\mathrm{eff}$ after the $\Lambda_\mathrm{eff}$ cut.
For both extrapolations we provide the uncertainty in the fit parameter $E_\infty$.

Unlike the ground-state energies, the convergence of rms radii with increasing $N_{\rm max}$ is not monotonic.  
Typically, for (moderately) large values of $\hbar\Omega$, the rms radii increase with increasing $N_{\rm max}$. For small $\hbar\Omega$ values (below the variational minimum in the ground state energies), the rms radii decrease with increasing $N_{\rm max}$.  
As a consequence, plots of the rms radii as function of the HO parameter $\hbar\Omega$ at fixed $N_{\rm max}$ show crossing points between curves for different $N_{\rm max}$ and often it appears as if all curves cross approximately at a single point at low $\hbar\Omega$.  In fact, it has been suggested \cite{Bogner:2007rx} that such a crossing point may be taken to provide a heuristic estimate for the true, converged value; however, such estimates are not necessarily found to be very robust~\cite{Caprio:2014iha}.  Nevertheless, we list the crossing points between successive $N_{\rm max}$ curves in \cref{tab:results} as a poor man's estimate of the converged values, without any uncertainty estimates.  Note that recently \cite{Caprio:2024tzt}, it has been shown that contrary to the na\"ive expectation, exponential extrapolation does seem to work for rms radii (and other long-range observables) when applied at moderately large values of $\hbar\Omega$, well above these crossing points.

\section{Discussion}

\begin{figure}
    \centering
    \includegraphics[width=\linewidth]{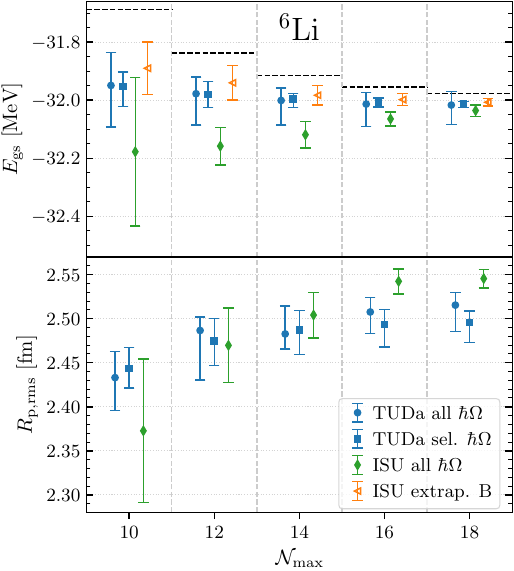}
    \caption{Extrapolations to the full Hilbert space for the ground-state energy and point-proton radius of $^6$Li based on NCSM calculations up to $\curly{N}_\mathrm{max}$ with the ISU ANNs (green) and the TUDa ANNs for the full frequency range $\hbar\Omega=10$ to $30$ MeV (blue circles) and for a selection of five optimal frequencies (blue squares). For the energy an additional exponential extrapolation (yellow) and the variational minima at the respective $N_\mathrm{max}$ (dashed lines) are given for comparison.}
    \label{fig:res_Li6}
\end{figure}

\begin{figure}
    \centering
    \includegraphics[width=\linewidth]{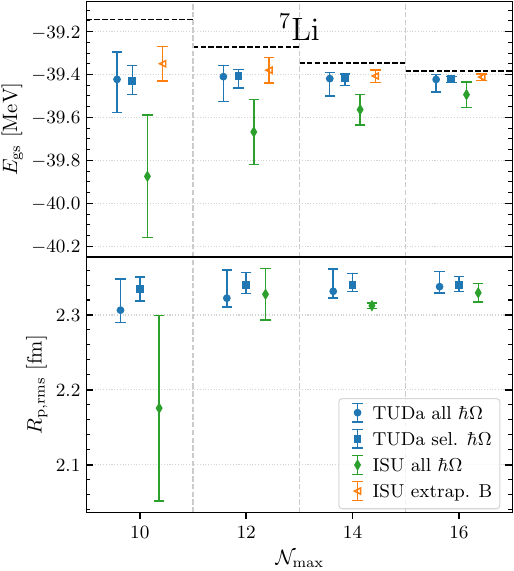}
    \caption{Same as \cref{fig:res_Li6} but for $^7$Li.}
    \label{fig:res_Li7}
\end{figure}

\begin{figure}[]
    \centering
    \includegraphics[width=\linewidth]{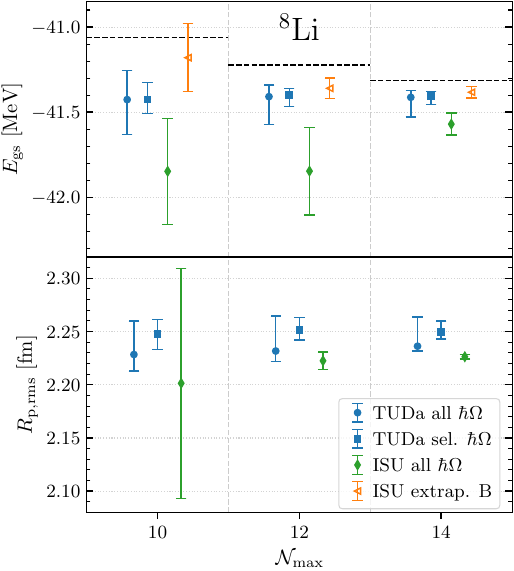}
    \caption{Same as \cref{fig:res_Li6} but for $^8$Li.}
    \label{fig:res_Li8}
\end{figure}

The extrapolations for the ground-state energies and point-proton radii of \elem{Li}{6}, \elem{Li}{7}, and \elem{Li}{8} based on the different ANNs, as well as the empirical exponential extrapolations for the energies are shown in \cref{fig:res_Li6,fig:res_Li7,fig:res_Li8}, respectively.
In addition, IR extrapolations for the energies and the crossing point estimates for the radii are given in \cref{tab:results}.  The predictions are grouped by $\curly{N}_\mathrm{max}$ indicating that the respective extrapolations are obtained based on data up to this model-space size. 
In regards to the quoted uncertainties, it should be noted that each method determined its uncertainty based on previously published studies.  The comparisons between the methods is the primary goal of the current work which, with future investigation, will lead to concrete estimates of systematic uncertainties.  We return to this point in the next section.

Starting with the ground-state energies depicted in the upper panels, we, first of all, find that neither the ANNs nor the exponential extrapolation violate the variational principle.  Note, however, that for the ISU ANNs this is part of the selection criteria for the best performing networks. 
In addition, the predictions from the different schemes exhibit different trends w.r.t. $\curly{N}_\mathrm{max}$.
The TUDa ANNs are very consistent across all $\curly{N}_\mathrm{max}$ and the predictions are significantly more precise for the selected $\hbar\Omega$ window. 
The ISU ANNs tend to overestimate the binding energy in small model spaces and show an upward trend with increasing $\curly{N}_\mathrm{max}$.
In contrast, the exponential extrapolation exhibits a downward trend.
For \elem{Li}{6} and \elem{Li}{7} all methods agree within uncertainties at the largest $\curly{N}_\mathrm{max}$ available, but at the smaller $\curly{N}_\mathrm{max}$ values the ISU ANN uncertainties do not overlap with either the selected TUDa ANN uncertainties or the ISU extrapolation B uncertainties; and also in \elem{Li}{8} the TUDa ANNs agree with the ISU exponential extrapolation, while the ISU ANNs predict a slightly lower energy, outside the uncertainties of the selected TUDa ANNs and those of the ISU exponential extrapolation. 

Considering the additional results in \cref{tab:results} we find that the na\"ive IR extrapolation violates the variational boundaries. This is remedied by the manual data selection. The improved extrapolations behave similarly to the exponential extrapolation, including the overall downward trend, with a tendency to larger binding energies (deeper bound).
Note that the uncertainties on the IR extrapolations do not include any uncertainties arising from the data selection procedure -- the quoted uncertainties are only the uncertainties in the fit parameter $E_\infty$.

\begin{table*}[]
    \centering
    \begin{ruledtabular}
    \begin{tabular}{l d{9} d{9} d{9} d{9} d{9}} 
    Method  &  \multicolumn{1}{c}{$\curly{N}_\text{max}=10$}  &  \multicolumn{1}{c}{12}  &  \multicolumn{1}{c}{14}  &  \multicolumn{1}{c}{16}  &  \multicolumn{1}{c}{18}\\
    \hline\\[-0.8em]
    \multicolumn{6}{l}{\hspace{12em}$^6$Li ground-state energy [MeV] \quad---\quad Experiment: \num{-31.993987 \,\pm\, 0.000002}} \\[.2em]
    \hline\\[-.8em]
    Variational minimum  &  -31.688  &  -31.837  &  -31.914  &  -31.954  &  -31.977 \\[.2em]%
    TUDa all $\hbar\Omega$  &  -31.949_{-0.143}^{+0.114}  &  -31.978_{-0.108}^{+0.057}  &  -32.001_{-0.085}^{+0.043}  &  -32.013_{-0.078}^{+0.040}  &  -32.017_{-0.067}^{+0.046}\\[.2em]%
    TUDa $\hbar\Omega$=\qtyrange{10}{20}{MeV}  &  -31.953_{-0.070}^{+0.050}  &  -31.980_{-0.044}^{+0.044}  &  -31.997_{-0.027}^{+0.021}  &  -32.009_{-0.013}^{+0.017}  &  -32.011_{-0.014}^{+0.006}\\[.2em]%
    ISU all $\hbar\Omega$  &  -32.18 \,\pm\, 0.26  &  -32.158 \,\pm\, 0.065  &  -32.119 \,\pm\, 0.045  &  -32.065 \,\pm\, 0.024  &  -32.036 \,\pm\, 0.020\\
    ISU extrap. B  &  -31.89 \,\pm\, 0.09  &  -31.94 \,\pm\, 0.06  &  -31.983 \,\pm\, 0.033  &  -31.998 \,\pm\, 0.021  &  -32.007 \,\pm\, 0.013\\
    IR extrap.  &  -31.437 \,\pm\, 0.050  &  -31.599 \,\pm\, 0.031  &  -31.742 \,\pm\, 0.019  &  -31.827 \,\pm\, 0.013  &  -31.888 \,\pm\, 0.009\\
    IR extrap. man. opt.  &  -31.72 \,\pm\, 0.14  &  -31.896 \,\pm\, 0.090  &  -31.967 \,\pm\, 0.080  &  -32.000 \,\pm\, 0.053  &  -31.969 \,\pm\, 0.037\\
    \hline\\[-0.8em]
    \multicolumn{6}{l}{\hspace{12em}$^6$Li point-proton rms radius [fm] \quad---\quad Experiment: \num{2.465 \,\pm\, 0.041}} \\[.2em]
    \hline\\[-0.8em]
    TUDa all $\hbar\Omega$  &  2.433_{-0.037}^{+0.029}  &  2.487_{-0.056}^{+0.015}  &  2.483_{-0.017}^{+0.032}  &  2.508_{-0.024}^{+0.017}  &  2.515_{-0.030}^{+0.014}\\[.2em]%
    TUDa $\hbar\Omega$=\qtyrange{15}{25}{MeV}  &  2.444_{-0.022}^{+0.024}  &  2.474_{-0.027}^{+0.026}  &  2.487_{-0.028}^{+0.023}  &  2.493_{-0.026}^{+0.017}  &  2.496_{-0.023}^{+0.013}\\[.2em]%
    ISU all $\hbar\Omega$  &  2.373 \,\pm\, 0.082  &  2.470 \,\pm\, 0.042  &  2.504 \,\pm\, 0.026  &  2.542 \,\pm\, 0.014  &  2.546 \,\pm\, 0.011\\
    Crossing point &  2.37  &  2.41  &  2.43  &  2.44  &   \\
    \hline\\[-0.8em]
    \multicolumn{6}{l}{\hspace{12em}$^7$Li ground-state energy [MeV] \quad---\quad Experiment: \num{-39.245081 \,\pm\, 0.000004}} \\[.2em]
    \hline\\[-0.8em]
    Variational minimum  &  -39.143  &  -39.271  &  -39.347  &  -39.383  \\[.2em]%
    TUDa all $\hbar\Omega$  &  -39.423_{-0.155}^{+0.128}  &  -39.410_{-0.116}^{+0.053}  &  -39.419_{-0.082}^{+0.030}  &  -39.423_{-0.058}^{+0.024}\\[.2em]%
    TUDa $\hbar\Omega$=\qtyrange{10}{20}{MeV}  &  -39.430_{-0.063}^{+0.072}  &  -39.409_{-0.053}^{+0.032}  &  -39.418_{-0.032}^{+0.019}  &  -39.419_{-0.018}^{+0.009}\\[.2em]%
    ISU all $\hbar\Omega$  &  -39.87 \,\pm\, 0.29  &  -39.67 \,\pm\, 0.15  &  -39.564 \,\pm\, 0.071  &  -39.494 \,\pm\, 0.059\\
    ISU extrap. B  &  -39.35 \,\pm\, 0.08  &  -39.38 \,\pm\, 0.06  &  -39.407 \,\pm\, 0.029  &  -39.412 \,\pm\, 0.015\\
    IR extrap.  &  -39.29 \,\pm\, 0.28  &  -39.270 \,\pm\, 0.099  &  -39.317 \,\pm\, 0.047  &  -39.352 \,\pm\, 0.024\\
    IR extrap. man. opt.  &  -39.36 \,\pm\, 0.19  &  -39.318 \,\pm\, 0.025  &  -39.377 \,\pm\, 0.020  &  -39.396 \,\pm\, 0.023\\
    \hline\\[-0.8em]
    \multicolumn{6}{l}{\hspace{12em}$^7$Li point-proton rms radius [fm] \quad---\quad Experiment: \num{2.321 \,\pm\, 0.044}} \\[.2em]
    \hline\\[-0.8em]
    TUDa all $\hbar\Omega$  &  2.306_{-0.016}^{+0.042}  &  2.322_{-0.012}^{+0.037}  &  2.332_{-0.009}^{+0.030}  &  2.338_{-0.008}^{+0.020}\\[.2em]%
    TUDa $\hbar\Omega$=\qtyrange{15}{25}{MeV}  &  2.335_{-0.017}^{+0.016}  &  2.340_{-0.011}^{+0.017}  &  2.340_{-0.009}^{+0.015}  &  2.340_{-0.009}^{+0.011}\\[.2em]%
    ISU all $\hbar\Omega$  &  2.18 \,\pm\, 0.12  &  2.328 \,\pm\, 0.034  &  2.312 \,\pm\, 0.004  &  2.330 \,\pm\, 0.012\\
    Crossing point &  2.28  &  2.30  &  2.31  &  2.32 \\
    \hline\\[-0.8em]
    \multicolumn{6}{l}{\hspace{12em}$^8$Li ground-state energy [MeV] \quad---\quad Experiment: \num{-41.277699 \,\pm\, 0.000047}} \\[.2em]
    \hline\\[-0.8em]
    Variational minimum  &  -41.062  &  -41.224  &  -41.313 \\[.2em]%
    TUDa all $\hbar\Omega$  &  -41.426_{-0.205}^{+0.170}  &  -41.409_{-0.164}^{+0.067}  &  -41.413_{-0.115}^{+0.039}\\[.2em]%
    TUDa $\hbar\Omega$=\qtyrange{10}{20}{MeV}  &  -41.426_{-0.081}^{+0.099}  &  -41.400_{-0.066}^{+0.040}  &  -41.403_{-0.051}^{+0.023}\\[.2em]%
    ISU all $\hbar\Omega$  &  -41.85 \,\pm\, 0.31  &  -41.85 \,\pm\, 0.26  &  -41.570 \,\pm\, 0.065 \\
    ISU extrap. B  &  -41.18 \,\pm\, 0.20  &  -41.36 \,\pm\, 0.06  &  -41.383 \,\pm\, 0.035\\
    IR extrap.  &  -40.785 \,\pm\, 0.085  &  -41.014 \,\pm\, 0.041  &  -41.199 \,\pm\, 0.020\\
    IR extrap. man. opt.  &  -41.31 \,\pm\, 0.21  &  -41.39 \,\pm\, 0.18  &  -41.38 \,\pm\, 0.11\\
    \hline\\[-0.8em]
    \multicolumn{6}{l}{\hspace{12em}$^8$Li point-proton rms radius [fm] \quad---\quad Experiment: \num{2.219 \,\pm\, 0.046}} \\[.2em]
    \hline\\[-0.8em]
    TUDa all $\hbar\Omega$  &  2.228_{-0.015}^{+0.031}  &  2.232_{-0.010}^{+0.033}  &  2.236_{-0.004}^{+0.028}\\[.2em]%
    TUDa $\hbar\Omega$=\qtyrange{15}{25}{MeV}  &  2.248_{-0.015}^{+0.014}  &  2.252_{-0.009}^{+0.011}  &  2.250_{-0.007}^{+0.010}\\[.2em]%
    ISU all $\hbar\Omega$  &  2.20 \,\pm\, 0.11  &  2.223 \,\pm\, 0.008  &  2.226 \,\pm\, 0.002\\
    Crossing point &  2.21  &  2.22  &  2.23
    \end{tabular}%
    \end{ruledtabular}%
    \caption{Extrapolations from different methods for the ground-state energy and point-proton rms radius of $^6$Li, $^7$Li, and $^8$Li using only information from NCSM calculations in model spaces $N_\text{max} \le \curly{N}_\text{max}$. The label ``all $\hbar\Omega$'' refers to $\hbar\Omega$=\qtyrange{10}{30}{MeV} with a stepsize of \qty{2.5}{MeV}. Experimental values are taken from Refs. \cite{Wang:2021xhn,Angeli:2013epw}.}
    \label{tab:results}
\end{table*}

Next, we move on to the predictions of radii, which are depicted in the lower panels of \cref{fig:res_Li6,fig:res_Li7,fig:res_Li8}.
Starting with \elem{Li}{6} we find an upward trend with $\curly{N}_\mathrm{max}$ for both ANN approaches. 
This trend is more pronounced for the ISU ANNs, which also predict a larger radius at $\curly{N}_\mathrm{max}=18$ than the TUDa ANNs with slightly smaller uncertainties.
Regarding consistency across different $\curly{N}_\mathrm{max}$, it appears that the TUDa ANNs are in agreement with each other within uncertainties starting from $\curly{N}_\mathrm{max}=12$, while for the ISU ANNs this only holds from $\curly{N}_\mathrm{max}=16$.
For \elem{Li}{7} we find both methods mostly in agreement with each other.
The TUDa ANN predictions are consistent across all $\curly{N}_\mathrm{max}$ and the frequency selection further improves the precision.
From $\curly{N}_\mathrm{max}=12$ we also find consistency for the ISU ANNs with the exception of a surprisingly large reduction of the uncertainty at $\curly{N}_\mathrm{max}=14$.
Lastly, in \elem{Li}{8} we find very stable predictions for both methods.
However, there is another large jump in precision for the ISU ANNs leading to a disagreement between the predictions at the largest $\curly{N}_\mathrm{max}$.
These unexpectedly precise radius predictions from the ISU ANNs at $\curly{N}_\mathrm{max} = 14$ for both \elem{Li}{7} and \elem{Li}{8} suggest that the assigned uncertainties are underestimated and do not resemble a 1-$\sigma$ interval.
Again, we can further compare this to the crossing point results given in \cref{tab:results}, which are in excellent agreement with the ISU ANNs for \elem{Li}{7} and \elem{Li}{8}, however, they significantly underestimate the predictions of both ANN methods for \elem{Li}{6}.

\section{Conclusions and Outlook}

In this work, we provide a detailed benchmark of various model-space extrapolation methods for ground-state energies and point-proton radii in Li isotopes. 
We find that modern ML extrapolation schemes can provide robust predictions for the converged values of these observables. 
They further allow for a statistical uncertainty quantification, which is crucial for precision studies of nuclear properties and has previously not been accessible for observables other than energies.

Our predictions for the ground-state energies are largely consistent with established exponential and IR extrapolations, and the ML radius predictions exceed the capabilities of the crossing point method.
In particular, the TUDa ANNs produce remarkably robust results in small model spaces already, in good agreement with the exponential extrapolation for the energies near the variational minimum, and thus providing a significant improvement over previous extrapolation schemes. 
Moreover, they are computationally efficient as the ANNs have to be trained only once and the comparably small number of NCSM calculations needed to obtain predictions allows for an application across the $p$-shell for any realistic Hamiltonian.

As already mentioned, the ISU ANNs tend to overestimate the binding energy
and give rather large uncertainties in small model spaces; and despite the large uncertainties, they often are not overlapping with the 'TUDa selected ANN' uncertainties, nor with the estimated uncertainties of the exponential extrapolation.
In addition, the ISU ANNs produce sometimes rather large jumps in either the predicted value or the associated uncertainties, in particular for the radii. 
This may be in part caused by the relatively small amount of training data, starting with only 45 input data points at $\curly{N}_\mathrm{max} = 10$.
Indeed, Refs.~\cite{Negoita:2018kgi,Negoita:2018yok} used about double the number of input data, with 19 different $\hbar\Omega$ values in the range from 8 to 50 MeV at most $N_\text{max}$ values. However, performing the NCSM calculations on such an extended range in $\hbar\Omega$ becomes computationally rather expensive, as is also mentioned in connection with the IR extrapolations.  Related to this is the fact that for the ISU approach we need input data up to at least $N_\text{max} = 10$, but preferably up to $N_\text{max} = 12$, before we have sufficient input data to train the ANNs.  This means that with a nucleon-nucleon potential we can apply this method for most $p$-shell nuclei, but once we include three-nucleon forces, this method is limited to only the lower half of the $p$-shell, at least with current high-performance computing resources.

In addition, our benchmarking process suggests that there may be systematic uncertainties in the ISU ANN procedures.  Indeed, the details of the selection procedure for the 50 best performing ANNs out of 200,000 trial networks (up to 120,000 trial networks in Ref.~\cite{Negoita:2018kgi}), such as the 'flatness criterion' at sufficiently large $N_{\rm max}$ or the monotonic decreasing behavior of the energies  with increasing $N_{\rm max}$, influence the ISU ANN predictions and associated uncertainties~\cite{McCarty:2024diss}.   While the current choice of the subset of ANNs to retain for the  histograms produces approximately internally consistent results~\cite{Negoita:2018kgi,McCarty:2024diss}, the selected subset of ANNs could be  enlarged to retain more networks with suitably larger cost function values in order to account for apparent systematic uncertainties.  Future works will investigate this line of thinking.

A final advantage of the discussed ML approaches is the extension to other observables.
The ISU ANNs can directly be applied to any observable, which has already been demonstrated for magnetic dipole and electric quadrupole moments~\cite{McCarty:2024diss}.
A direct application of the TUDa ANNs to these observables is challenging due to limited training data, however, this framework has recently been extended to describe electric quadrupole moments \cite{Knoell:2025otn}.

\section*{Acknowledgments}

This work is supported by the Deutsche Forschungsgemeinschaft (DFG, German Research Foundation) through the DFG Sonderforschungsbereich SFB 1245 (Project ID 279384907), 
the BMBF through Verbundprojekt 05P2024 (ErUM-FSP T07, Contract No. 05P24RDB), and by the U.S. Department of Energy, Office of Science, under Award Numbers DE-SC0023495 (SciDAC5/NUCLEI) and DE-SC0023692.
The NCSM calculations for the results presented here were performed at the National Energy Research Scientific Computing Center (NERSC), a DOE Office of Science User Facility supported by the Office of Science of the U.S. Department of Energy under Contract No. DE-AC02-05CH11231, using NERSC awards NP-ERCAP0020944 (2022), NP-ERCAP0023866 (2023), and NP-ERCAP0028672 (2024).
Numerical calculations of the training data for the TUDa ANNs have been performed on the LICHTENBERG II cluster at the computing center of the TU Darmstadt.

\bibliographystyle{apsrev4-2}
\bibliography{references.bib}

\end{document}